\documentclass[a4paper,fleqn]{cas-dc}

\usepackage[numbers]{natbib}
\usepackage{threeparttable}

\def\tsc#1{\csdef{#1}{\textsc{\lowercase{#1}}\xspace}}
\tsc{WGM}
\tsc{QE}
\tsc{EP}
\tsc{PMS}
\tsc{BEC}
\tsc{DE}

\usepackage{enumitem}
\usepackage{hyperref}
\usepackage{ulem}
\usepackage{float}

\def\be{\begin{eqnarray}}
\def\ee{\end{eqnarray}}

\begin{document}
\let\WriteBookmarks\relax
\def\floatpagepagefraction{1}
\def\textpagefraction{.001}
\shorttitle{Scale symmetry in nuclear medium}
\shortauthors{Y.L. Ma}

\title [mode = title]{Manifestation of scale symmetry in nuclear medium}                      
\tnotemark[1]

\tnotetext[1]{Contribution to the volume in memorial of Prof. Mannque Rho.}


\author[1]{Yong-Liang Ma}[type=editor,
                        auid=000,bioid=1,
                        orcid=0000-0001-9154-529X]
\ead{ylma@nju.edu.cn}


\affiliation[1]{organization={School of Physics, Nanjing University},
                city={Nanjing},
                postcode={210093}, 
                state={Jiangsu},
                country={China}}









\begin{abstract}
This memorial tribute to Mannque Rho focuses on the role of scale symmetry in baryonic matter, especially on its profound influence on the possible resolution of the ``quenched $g_A$" problem in the Gamow-Teller transition of nuclei, the pseudoconformal structure of dense nuclear matter and the origin of the peak of sound velocity of nuclear matter. The discussion is based on the genuine dilaton scenario, anchored on the Crewther-Tunstall infrared fixed point, which provides a unified framework where scale symmetry—broken explicitly by the QCD trace anomaly—can be spontaneously broken, yielding the scalar $f_0(500)$ as a pseudo-Nambu-Goldstone boson. In nuclear medium, this scenario serves as a natural foundation of the Brown-Rho scaling---a relation that has been central to nuclear physics for over three decades. The sound velocity of nuclear matter exhibits a characteristic peak at intermediate densities arising from the kink behavior of the dilaton field, a unique signature of scale symmetry realization that distinguishes the dilaton-compensated effective theory from conventional Walecka-type models. At higher densities, the approach to the dilaton-limit fixed point produces pseudo-conformal behavior with sound velocity $v_s^2 = 1/3$ saturating from below at $n \gtrsim 3n_0$, while the trace of the energy-momentum tensor remains nonzero. The framework also provides a natural resolution of the long-standing ``quenched $g_A$" problem, demonstrating that nuclear correlations, not fundamental quenching, account for the effective axial coupling approaching unity. The continuity from the unitary limit at low density to the dilaton limit at high density suggests a pervasive role for scale symmetry in nuclear dynamics across all density regimes.
\end{abstract}




\maketitle

\section{Introduction}

It is well accepted and established that chiral symmetry plays dominant roles in hadronic processes. However, another fundamental symmetry of QCD, the scale symmetry which exists in QCD Lagrangian in chiral limit but is explicitly broken by quantum corrections~\cite{Coleman:1985rnk}, has not received full attention in hadron and nuclear physics. Although there are several discussions in literature~\cite{Isham:1970xz,Ellis:1970yd,Schechter:1980ak,Migdal:1982jp,Campbell:1989gh,Campbell:1990ak}, to my opinion, the pivotal development in this context is attributed to Gerry Brown and Mannque Rho in 1991~\cite{Brown:1991kk}. This memorial contribution is devoted to some effects of scale symmetry on nuclear dynamics.

In the pioneering work~\cite{Brown:1991kk}, by using the dilaton-compensated Skyrme model, Brown and Rho proposed that the nucleon mass, pion decay constant and vector meson masses have approximately the same density dependence that is attributed to the vacuum expectation value (VEV) of dilaton field, the so-called Brown-Rho (BR) scaling relation.  Since then, the BR scaling relation has been central to nuclear physics for over three decades and, the manifestation/emergence of scale symmetry in (dense) nuclear matter (NM) has drown more attention (see, e.g., Refs.~\cite{Brown:1995qt,Brown:2001nh,Lee:2003eg,Park:2008zg,Holt:2014hma,Ma:2019ery} and references therein and other contributions to this volume).

In the chiral effective field theories (EFTs) which are anchored in the nonlinear realization of chiral symmetry of QCD and have been widely and successfully used in hadron and nuclear physics~\cite{Weinberg:1990rz,Weinberg:1991um,Machleidt:2011zz,Holt:2014hma}, the scalar meson field---the fourth component of chiral four-vector---is integrated out when going from the linear realization to the nonlinear realization of chiral symmetry. Therefore, the iso-scalar scalar meson $\sigma$---an important degree of freedom (DoF) responsible for the attractive force between nucleons---is absent. This limits the applicability of chiral EFTs in nuclear physics, particularly in the relativistic mean-field (RMF) approach~\cite{Holt:2014hma}. A way out of this limitation is to consider the contact terms of nucleon interaction~\cite{Weinberg:1990rz,Weinberg:1991um,Machleidt:2011zz} which introduce more parameters in the theory, or to introduce a scalar field based on certain symmetry of QCD that will be focused on in this contribution.

To incorporate  the scalar mesons in hadronic models, a certain symmetry of QCD which plays the role of the source is necessary. The trace anomaly of QCD takes the responsibility of the source of the iso-scalar scalar meson with quantum numbers $J^{PC}=0^{++}$~\cite{Schechter:1980ak}. In order to construct the EFT of scalar meson with consistent power counting, Crewther and Tunstall  (hereafter referred to as CT) proposed~a genuine dilaton approach to include the $\sigma$ meson---can be identified with $f_0(500)$ in the Particle Data Booklet---in chiral EFT~\cite{Crewther:2013vea,Cata:2018wzl,Crewther:2020tgd} by conjecturing that QCD has an infrared fixed point (IRFP) and that the theory is slightly away from it. In this approach, the $\sigma$ meson is identified as the Nambu-Goldstone boson (NGB) of scale symmetry breaking---the dilaton---with its small mass arising from the deviation from the IRFP that explicitly breaks scale symmetry. The existence of the IRFP is supported by some literature, e.g., Refs.~\cite{Cui:2019dwv,Deur:2025rjo}. 

In this contribution, I will discuss some features of NM which are difficult---if not impossible---to be predicted by other approaches by using the dilaton compensated baryonic hidden local symmetry (b$s$HLS)---with ``b"  standing for baryon, ``$s$" for scale symmetry and HLS for hidden local flavor symmetry---which we developed over last decade~\cite{Li:2016uzn,Paeng:2017qvp}. The results trace the evolution of scale symmetry from dilute nuclear matter to the ultra-dense regime of compact stars.

Explicitly, I will first briefly describe the features of b$s$HLS and the meaning of BR scaling in Sec.~\ref{sec:bsHLS}. This illustrates the theoretical foundations of scale symmetry in QCD and the genuine dilaton scenario, and how BR scaling emerges as a natural consequence. Then, I will discuss how the b$s$HLS can give a resolution of the ``quenched $g_A$ problem" in weak transitions of nuclei in Sec.~\ref{sec:gA} to show how the scale symmetry manifests in nuclei. In Sec.~\ref{sec:PCM}, the pseudoconformal structure---the sound velocity (SV) saturating the conformal limit $v_s^2=1/3$ but the trace of energy momentum tensor (TEMT) does not vanish---of dense NM are predicted. To my best knowledge, this structure was not observed before us. Due to the dilaton compensator approach, the peak of SV emerges naturally in the b$s$HLS, without resort to the hadron-quark transition or the emergence of new DoFs. This aspect will be illustrated in Sec.~\ref{sec:peak}. My summary and perspective are finally presented in Sec.~\ref{sec:sum}.

\section{Dilaton compensated effective field theories and Brown-Rho scaling}

\label{sec:bsHLS}

\subsection{Scale Symmetry in QCD and the Genuine Dilaton Scenario}

In chiral limit, the QCD Lagrangian is classically invariant under the scale transformation of coordinates
\be
x_\mu \to \lambda^{-1} x_\mu,
\label{eq:defScale}
\ee
with $\lambda$ being a constant. Therefore, the TEMT $\theta_\mu^\mu$ vanishes at classical level, i.e., $ \theta_\mu^\mu =0$, and the dilatation current $D^\mu$ is conserved $\partial_\mu D^\mu = 0$. However, quantum effects introduce the trace anomaly:
\be
\theta_\mu^\mu  =   \frac{\beta(\alpha_s)}{4\alpha_s} G_{\mu\nu}^a G^{a\mu\nu} + \left(1+\gamma_m(\alpha_s)\right)\sum_q m_q \bar{q}q,
\label{eq:TAQCD}
\ee
where the quark mass term is included, $\beta(\alpha_s)$ is the QCD beta function and $\gamma_m(\alpha_s)$ is the anomalous dimension of the quark mass operator. The first term, proportional to the gluon condensate in vacuum, represents the explicit breaking of scale symmetry that persists even when quark masses vanish.

Trace anomaly~\eqref{eq:TAQCD} is written in terms of quarks and gluons, the DoFs in QCD. When matching to the low energy process of strong interaction where the colors are confined, This effect should be accounted for in terms of a colorless object $\chi$ through the anomaly matching condition:
\be
\theta_\mu^\mu \Big|_{\rm QCD} = \frac{\beta(\alpha_s)}{4\alpha_s} G_{\mu\nu}^a G^{a\mu\nu}   = \theta_\mu^\mu \Big|_{\rm EFT}  \propto  \chi^4,
\label{eq:TAmatch}
\ee
where the dilaton compensator field $\chi$ has scale dimension-1 and the coefficient in front of $\chi^4$ will be given later. Note that the dilaton field $\chi$ only mimics the gluon part of trace anomaly and the matching condition \eqref{eq:TAmatch} only involves the gluon part of~\eqref{eq:TAQCD}.  Since $\theta_\mu^\mu$ has the quantum numbers of vacuum, the fluctuation of $\chi$ with respect to its VEV $\langle \chi \rangle \equiv f_\chi$ can be regarded as the iso-scalar scalar meson in hadronic models~\cite{Isham:1970xz,Ellis:1970yd,Schechter:1980ak,Crewther:2013vea,Cata:2018wzl}, explicitly, the lightest scalar meson $f_0(500)$ in Particle Data Booklet.  Actually, the anomaly-matching equation~\eqref{eq:TAmatch} is the only constraint on the appearance of $\chi$ in effective models. Any functionals of $\chi$ satisfying this constraint from symmetry argument in the general effective models are permitted. The other constraints come from the physical reality.

If one were only considering \eqref{eq:TAmatch}, it is impossible to develop an EFT of the lightest scalar meson $f_0(500)$ \'a la Weinberg with a self-consistent power counting mechanism as in the chiral EFT. A resolution was proposed by Crewther and Tunstall who posit that QCD has an IRFP at $\alpha_{\rm IR}$ where $\beta(\alpha_{\rm IR})=0$, allowing scale symmetry to be spontaneously broken with  the Nambu-Goldstone boson being  dilaton---the genuine dilaton~\cite{Crewther:2013vea,Cata:2018wzl,Crewther:2020tgd}. The departure of the EFT from the fixed point, parameterized by $\beta^\prime = \partial \beta/\partial \alpha_s|_{\alpha_{\rm IR}}$which is related to the anomalous dimension of gluon field $\gamma_{G^2}$ at a (presumed) IFFP through $\gamma_{G^2}=-\beta^\prime$~\cite{Ryttov:2016hal}, provides the explicit breaking  that generates the dilaton mass in chiral limit:
\be
m_\chi^2f_\chi^2 =  \frac{\beta^\prime \Delta\alpha_s}{\alpha_{\rm IR}} \left\langle G_{\mu\nu}^2 \right\rangle ,
\label{eq:GMORs}
\ee
where $\Delta\alpha_s = \alpha_{\rm IR} - \alpha_s$ with $\alpha_s$ being the strong coupling constant where the EFT is built. Therefore, the magnitude of dilaton mass is proportional to the deviation from the IRFP, $\Delta \alpha = \alpha_{IR}-\alpha_s$ accounting for the magnitude of the explicit scale symmetry breaking.  Eq.~\eqref{eq:GMORs} is the Gell-Mann--Oakes--Renner (GMOR) type relation in genuine dilaton sector. As a result, the situation is very similar to the chiral perturbation theory where the masses of the Nambu-Goldstone bosons are proportional to the current quark masses which measure the magnitude of the explicit chiral symmetry breaking. Note that whether the nonperturbative IRFP which the CT approach is anchored on exists in QCD is not yet confirmed. The arguments given in, e.g., Refs.~\cite{Cui:2019dwv,Deur:2025rjo} support this proposition.

\subsection{Dilaton compensated effective theories}

Following CT~\cite{Crewther:2013vea,Cata:2018wzl}, introducing the conformal compensator field $\chi$ which has the scale dimension 1 and $\langle\theta_\mu^\mu\rangle \propto \langle \chi^4\rangle = f_\chi^4$, one can write the chiral-scale effective Lagrangian in terms of the scale dimensions in the chiral-scale power counting in momentum (derivative), quark mass and $\Delta\alpha=\alpha_{IR}-\alpha_s$
\be
\mathcal{O}(p^2) \sim \mathcal{O}(m_q) \sim \mathcal{O}(\Delta\alpha),
\ee
where the last counting is from the relation~\eqref{eq:GMORs}.

For simplicity, let us first consider the pseudoscalar pions $U(x) = e^{2i\pi(x)/f_\pi}$ and dilaton $\chi = f_\chi e^{\sigma(x)/f_\chi}$ with $\sigma$ being the dilaton field which will be identified with the lightest scalar meson $f_0(500)$. The effective Lagrangian at leading chiral-scale order can be decomposed as
\be
{\cal L}_{\rm LO} & = & {\cal L}_{\rm inv}^{d=4} + {\cal L}_{\rm anom}^{d > 4} + {\cal L}_{\rm mass}^{d < 4}
\label{eq:LOChiralS}
\ee
where $d$ stands for the scale dimension and
\begin{subequations}
	\begin{eqnarray}
		{\cal L}_{\rm inv}^{d=4} & = & c_1 \frac{f_\pi^2}{4} \left( \frac{\chi}{f_\chi}\right)^2 {\rm Tr}\left( \partial_\mu U \partial^\mu U^\dagger \right) + \frac{1}{2} c_2 \partial_\mu \chi \partial^\mu \chi \nonumber\\
		& &{} + c_3 \left( \frac{\chi}{f_\chi}\right)^4, \label{eq:CTL40}\\
		{\cal L}_{\rm anom}^{d > 4} & = & (1 - c_1)\frac{f_\pi^2}{4} \left( \frac{\chi}{f_\chi}\right)^{2+\beta^\prime} {\rm Tr}\left( \partial_\mu U \partial^\mu U^\dagger \right) \nonumber\\
		& &{} + \frac{1}{2}(1 - c_2) \left( \frac{\chi}{f_\chi}\right)^{\beta^\prime} \partial_\mu \chi \partial^\mu \chi \nonumber\\
		& &{} + c_4 \left( \frac{\chi}{f_\chi}\right)^{4+\beta^\prime},\label{eq:CTLg40}\\
		{\cal L}_{\rm mass}^{d < 4} & = &{} \frac{f_\pi^2}{4} \left( \frac{\chi}{f_\chi}\right)^{3-\gamma_m} {\rm Tr}\left( \mathcal{M}^\dagger U + U^\dagger \mathcal{M} \right),\label{eq:CTLm40}
	\end{eqnarray}
\end{subequations}
where $\mathcal{M}$ stands for the current quark matrix with $\mathcal{M} = {\rm diag}(m_\pi^2,m_\pi^2)$ and $c_i$'s are unknown constants. Here, for the present purpose we only consider two flavors and remove the strange degree of freedom from the original CT approach. It should be noted that, different from the chiral perturbation theory, $c_3$ and $c_4$ have scale-chiral order $O(p^2)$ since, as will be seen later, they are proportional to the dilaton mass square, similarly to $\mathcal{M}$ for the pseudo-scalar Nambu-Goldstone mesons.

We next consider the dilaton potential $V(\chi)$
\be
V(\chi)  = {} - c_3 \left( \frac{\chi}{f_\chi}\right)^4 - c_4 \left( \frac{\chi}{f_\chi}\right)^{4+\beta^\prime}.
\ee
By using the saddle point equation and definition of the dilaton mass $m_\sigma$, one obtains
\be
V(\chi)  =  {} \frac{m_\sigma^2f_\chi^2}{4\beta^\prime(4+\beta^\prime)}\left( \frac{\chi}{f_\chi}\right)^4\left[- (4 + \beta^\prime) + 4 \left( \frac{\chi}{f_\chi}\right)^{\beta^\prime}\right].\nonumber\\
\ee
When $\beta^\prime \ll 1$ which works for some models of electroweak symmetry breaking, the dilaton potential is approximated to~\cite{Goldberger:2007zk,Matsuzaki:2013eva}
\begin{equation}
	V(\chi) = \frac{m_\sigma^2 f_\chi^2}{4} \left(\frac{\chi}{f_\chi}\right)^4 \left(\ln \frac{\chi}{f_\chi} -\frac{1}{4}\right).
\label{anomaly1}
\end{equation}
However, in QCD, we do not have any prior information on $\beta^\prime$. Its magnitude is estimated as $1.0\lesssim \beta^\prime \lesssim 3.5$ using the skyrmion crystal approach~\cite{Ma:2016nki,Shao:2022njr}. This yields the scale Ward-Takahashi identity
\be
\langle\theta_\mu^\mu\rangle=\langle\partial_\mu D^\mu\rangle ={} - \frac{m_\sigma^2f_\chi^2}{4+\beta^\prime}\left\langle\left(\frac{\chi}{f_\chi}\right)^{4+\beta^\prime}\right\rangle
\ee
which results to
\be
\langle\theta_\mu^\mu\rangle=\langle\partial_\mu D^\mu\rangle ={} - \frac{m_\sigma^2}{4f_\chi^2}\left\langle\chi^4\right\rangle
\ee
for $\beta^\prime \ll 1$. These two equations are the partially conserved dilatonic current (PCDC) relations, the counterpart to the PCAC for the pion.

Along the reasoning of CT, one can set up a systematic higher-order expansion and write down the higher order terms~\cite{Li:2016uzn,Cata:2019edh}. This is beyond the scope of this contribution, we will not go to the details.

In the general Lagrangian~\eqref{eq:LOChiralS}, there are so many unknown parameters that it is difficult to give any prediction in practice. One can arrive at a manageable form by taking the so called ``leading-order scale symmetry (LOSS)'' approximation that corresponds to
\be
c_1\approx c_2\approx 1.
\label{eq:LOSSCi}
\ee
That is, in LOSS, the scale symmetry breaking---in the chiral limit---is lodged entirely in the dilaton potential $V(\chi)$. The resulting Lagrangian is
\be
	{\cal L}_{{\rm LO}}^{\chi{\rm limit}} & = & \frac{f_\pi^2}{4} \left( \frac{\chi}{f_\chi}\right)^2 {\rm Tr}\left( \partial_\mu U \partial^\mu U^\dagger \right) \nonumber\\
	& &{} + \frac{1}{2} \partial_\mu \chi \partial^\mu \chi
	- V(\chi).
	\label{LOSS}
\ee
Whether the LOSS approximation is valid cannot be justified from first principles. The numerical analysis shows that it works for light nuclei~\cite{Li:2017udr,Li:2018ykx} and compact star matter~\cite{Ma:2019ery} but it violates around saturation density~\cite{Ma:2020tsj}.

The above procedure can be easily extended to include other hadron degrees of freedom, e.g.,  baryons as matter fields and the lowest-lying vector mesons $\rho$ and $\omega$ through the HLS approach~\cite{Harada:2003jx}. We decompose the Lagrangian into the pure meson part \(\mathcal{L}_M\) and the baryon part \(\mathcal{L}_B\) as~\cite{Zhang:2024sju,Zhang:2024iye}
\begin{equation}
	\mathcal{L}=\mathcal{L}_M+\mathcal{L}_B ,
	\label{eq:LbsHLS}
\end{equation}
where, in terms of $\Phi \equiv \chi/f_\chi$, we express
\be
\mathcal{L}_M & = & \ f^2_\pi\Phi^2{\rm Tr}\left(\hat{\alpha}^\mu_\perp\hat{\alpha}_{\mu\perp}\right) +\frac{m^2_\rho}{g^2_\rho}\Phi^2{\rm Tr}\left(\hat{\alpha}^\mu_\parallel\hat{\alpha}_{\mu\parallel}\right)\nonumber \\
& &{} +\frac{1}{2}\left(\frac{m^2_\omega}{g^2_\omega}-\frac{m^2_\rho}{g^2_\rho}\right)\Phi^2{\rm Tr}\left(\hat{\alpha}^\mu_\parallel\right){\rm Tr}\left(\hat{\alpha}_{\mu\parallel}\right)\nonumber\\
& &{} -\frac{1}{2g^2_\rho}{\rm Tr}\left(V_{\mu\nu}V^{\mu\nu}\right) -\frac{1}{2g^2_0}{\rm Tr}\left(V_{\mu\nu}\right){\rm Tr}\left(V^{\mu\nu}\right) \nonumber\\
& &{} +\frac{1}{2}\partial_\mu\chi\partial^\mu\chi+\frac{f^2_\pi}{4}\Phi^{3-\gamma_m}{\rm Tr}\left(\mathcal{M}U^\dagger+U\mathcal{M}^\dagger\right)\nonumber\\
& &{} +h_5\Phi^4+h_6\Phi^{4+\beta'}, \label{eq:LM}\\
\mathcal{L}_B & = & \bar{N} i \gamma_\mu D^\mu N-m_N \Phi \bar{N} N \nonumber\\
& &{} +\left[g_A C_A+g_A\left(1-C_A \right)\Phi^{\beta^{\prime}}\right] \bar{N} \hat{\alpha}_{\perp}^\mu \gamma_\mu \gamma_5 N \nonumber\\
& &{} +\left[g_{V_\rho} C_{V_\rho}+g_{V_\rho}\left(1-C_{V_\rho}\right) \Phi^{\beta^{\prime}}\right] \bar{N} \hat{\alpha}_{\|}^\mu \gamma_\mu N \nonumber\\
& &{} +\frac{1}{2}\left[g_{V_0} C_{V_0}+g_{V_0}\left(1-C_{V_0}\right) \Phi^{\beta^{\prime}}\right]\nonumber\\
& &{}\qquad \times {\rm Tr}\left[\hat{\alpha}_{\|}^\mu\right] \bar{N} \gamma_\mu N . 
\label{eq:LB}
\ee
Here, the two $1$-forms are
\begin{eqnarray}
	\hat{\alpha}_{\parallel\mu} & = & \frac{1}{2i}(D_\mu \xi_R \cdot
	\xi_R^\dag +
	D_\mu \xi_L \cdot \xi_L^\dag), \nonumber\\
	\hat{\alpha}_{\perp\mu} & = & \frac{1}{2i}(D_\mu \xi_R \cdot
	\xi_R^\dag - D_\mu \xi_L \cdot \xi_L^\dag)
	\ , \label{eq:1form}
\end{eqnarray}
where the covariant derivative is defined as $D_\mu \xi_{R,L} = (\partial_\mu - i V_\mu)\xi_{R,L}$ with
\begin{eqnarray}
	\xi_L^\dag & = & \xi_R \equiv \xi = e^{i\pi/(2f_\pi)}, \;\;\;\; U(x)
	= \xi^2(x). \label{eq:hiddenbreaking}
\end{eqnarray}
For the gauge field $V_\mu= \frac{1}{2}\left(g_\rho \rho_\mu^a \tau^a + g_\omega \omega_\mu\right)$, we have the field strength tensor
\begin{eqnarray}
	V_{\mu\nu}(x) = \partial_\mu V_\nu(x)-\partial_\nu
	V_\mu(x)-i[V_\mu(x),V_\nu(x)].
	\label{eq:hiddentensor}
\end{eqnarray}

In \eqref{eq:LbsHLS}, we considered the LOSS in the mesonic part. But in the baryonic part, some corrections to LOSS are included with $C_i \ne 1$ for $i=A, V_\rho, V_0$. The latter are found crucial for understanding the quenched $g_A$ in the weak decay of heavy nuclei~\cite{Ma:2020tsj}. For the details of theory construction, we referred to~\cite{Li:2016uzn,Zhang:2024sju,Zhang:2024iye}.

Since the b$s$HLS Lagrangian~\eqref{eq:LbsHLS} includes, in addition to the Nambu-Goldstone bosons pions and $\sigma$-dilaton,  the hadron resonances $\rho$ and $\omega$, the obtained equation of state (EoS) is expected to be applicable to the cores of massive stars, i.e., $\sim 10 n_0$.

\subsection{Brown-Rho scaling as a manifestation of scale symmetry}

Now, we are ready to discuss the phenomenological Brown-Rho scaling relation, proposed in 1991 whose theoretical foundation can now be taken as the scale-symmetric effective theory. Its validity, or the agreement of its prediction with experimental measurements, can be regarded as a manifestation of scale symmetry in hadron and nuclear physics.

The basic assumption of BR scaling is the following~\cite{Brown:1991kk}: In the dense medium, the symmetries of the Lagrangian are more or less intact with the variation of density and, the change of the vacuum which is reflected in the modifications of the quark and gluon condensates in QCD variables is encoded in the modified Lagrangian parameters, such as the masses and coupling constants according to the symmetry constraints of QCD. This assumption is not difficult to understand in the sense of the matching between Green function generating functionals of QCD and effective theories~\cite{Gasser:1983yg,Gasser:1987rb,Wang:1999cp}. This density scaling is attributed to the variation of dilaton condensate in medium
\be
\langle 0^\ast|\chi|0^\ast \rangle =  \langle \chi \rangle^\ast
\ee
with $|0^\ast \rangle$ being the ground state in medium.

Then, by coupling dilaton to the Skyrme model and encode trace anomaly into the dilaton potential (LOSS in the updated terminology), it is found that the effective pion decay constant is
\be
\frac{f_\pi^\ast}{f_\pi} = \frac{\langle \chi \rangle^\ast}{\langle \chi \rangle} \equiv \Phi^\ast(n).
\ee
While the pion mass remains unscaled $m_\pi^\ast/m_\pi \approx 1$. This reflects the near-Nambu-Goldstone nature of the pion. Since in the Skyrme model, the nucleon mass---skyrmion mass---depends on pion decay constant $m_N \propto f_\pi$, it is concluded that
\be
\frac{m_N^\ast}{m_N} = \frac{f_\pi^\ast}{f_\pi} = \Phi^\ast(n).
\ee
The density scaling of the vector meson masses can be derived from the Kawarabayashi-Suzuki-Riazuddin-Fayyazuddin (KSRF) relation
\be
m_V^2 = 2g^2f_\pi^2,
\label{eq:KSRF}
\ee
where $g$ is the hidden gauge coupling constant. Since in the Skyrme model, $g \propto \epsilon^{-1}$ which is scale invariant, one obtains
\be
\frac{m_V^\ast}{m_V} = \frac{f_\pi^\ast}{f_\pi} =  \Phi^\ast(n).
\ee
where, when the $O(1/N_c)$ correction is ignored, $V = \rho, \omega $. Furthermore, when the fluctuation of dilaton with respect to the vacuum is considered $\sigma \to \sigma_0^\ast + \tilde{\sigma}$, that is
\be
\chi = \chi^\ast \tilde{\chi} ,
\ee  
the effective dilaton mass---scalar meson mass---scales as
\be
\frac{m_\sigma^\ast}{m_\sigma} =  \Phi^\ast(n)
\ee

The above discussed CT scale-chiral EFT provides  a natural foundation of the BR scaling. Within the genuine dilaton scenario, BR scaling emerges naturally from the coupling of parameters to the dilaton condensate. The conformal compensator field $\chi $ enters the Lagrangian through terms such as:
\be
m_N \frac{\chi}{f_\chi}\bar{N}N, ~~ m_V^2 \frac{\chi^2}{f_\chi^2}V_\mu V^\mu, ~~ \frac{f_\pi^2}{4} \frac{\chi^2}{f_\chi^2}{\rm Tr}\left(\partial_\mu U \partial^\mu U^\dagger\right).
\ee
Therefore, the above discussed scaling relations obtained from Skyrme model follow directly.

Considering the corrections to the LOSS and sizable $\beta^\prime$ in QCD, the scaling relations are modified as 
\be
\frac{m_N^\ast}{m_N} \approx \frac{m_V^\ast}{m_V} \approx\frac{f_\pi^\ast}{f_\pi} \approx \frac{f_\chi^\ast}{f_\chi} \approx  \Phi^\ast(n),
\label{eq:BRScaling}
\ee
with the important distinction that the dilaton mass scales as 
\be
\frac{m_\sigma^\ast}{m_\sigma} = \Phi^\ast(n)^{1+\beta^\prime/2},
\label{eq:BRScalingsigma}
\ee 
reflecting the explicit breaking of scale symmetry encoded in the anomalous dimension $\beta^\prime$.

\subsection{Parameterization of Brown-Rho scaling}
\label{subsec:ParaBR}

The BR scaling gives a possible understanding of the origin of the medium modified bare parameters in the Lagrangian. How they scale explicitly with density cannot be determined from the first principle of strong interaction, so that some phenomenological information should be borrowed. 

It is recognized that hadron masses decrease with density, at least up to around saturation density $n_0$~\cite{Brown:1988de}. At higher densities $n \gtrsim 2.0n_0$, there is no empirical information as these densities are beyond the ability of the terrestrial nuclear physics experiments. The skyrmion crystal approach to NM found that the pion decay constant and nucleon mass first decrease with density at low densities, in agreement with the empirical information, and then stay more or less as a constant at higher densities, delineated by the topology change density $n_{1/2}$~\cite{Lee:2003aq,Ma:2013ooa}---a robust conclusion in the skyrmion matter.

Then, the density dependence of the medium modified hadron properties are categorized into two regions delineated by $n_{1/2}$. We denote the region $n<n_{1/2}$ as R-I and region $n>n_{1/2}$ as R-II. The density scaling of the medium modified hadron properties are summarized as follows:
\begin{itemize}
	\item R-I: In this region, the scaling function $\Phi^\ast$ decreases with density. Without first principle information, we parameterize it as
	\be
	\Phi^\ast_I=\frac{1}{1+r\frac{n}{n_0}}
	\label{eq:BRSRI}
	\ee
	with $r$ being a constant. With respect to the properties of pion atoms~\cite{Kienle:2004hq,piAF:2022gvw}, the range of $r$ is found to be
	\be
	r\approx 0.13-0.20 . 
	\label{eq:CIrange}
	\ee

	\item R-II: In this region, the scaling behaviours of the parameters are quite involved. What we know is that they are weaker than in R-I. They are summarized as follows:
	\begin{itemize}
		\item $g_\rho$ and $\rho$ mass: The hidden local gauge coupling $g_\rho$ relates to the $\rho$ mass through the KSRF relation~\eqref{eq:KSRF}. Combined with the vector manifestation (VM) fixed-point structure of HLS~\cite{Harada:2000kb,Harada:2001rf} leads to that the coupling $g_\rho$ should drop to zero at the putative VM fixed-point density $n_{VM}$. We take the simple form~\cite{Paeng:2017qvp}
		{\small
		\be
		\frac{g^*_{\rho NN}}{{g_{\rho NN}}}=\begin{cases}
			1-0.1\frac{n}{n_0}, &{\rm for~} n\in (n_{1/2}, 3.5 n_0) \\
			0.65-0.04\frac{(n-3.5n_0)}{n_0}, &{\rm for~} n\in (3.5n_0, n_{VM})
		\end{cases} \nonumber
		\ee
	}
		with $n_{\rm VM} \approx 20n_0$, beyond the density of massive compact stars.  Note that we do not have any priori information about the value of $n_{VM}$ but it is expected to be the region where the perturbative QCD is applicable.  Although the present choice predicts a pseudoconformal structure of neutron star matter, a smaller value even in the cores of massive neutron stars is acceptable numerically since it cannot be ruled out by the present measurements of the global properties of neutron stars. 
		
		\item $\omega$ meson: 
		Since the $U(2)$ HLS which works well in R-I breaks in R-II~\cite{Paeng:2011hy,Paeng:2015noa}, some sort of fine-tuning is needed in the density scaling of $\omega$ mass and hidden gauge coupling constant, which makes them different from that of rho meson. Since the NM properties are sensitive the omega repulsive force, the density scaling of $g_\omega$ and $m_\omega^\ast$ are both highly constrained and should weakly depends on density~\cite{Paeng:2015noa}. We take them as
		\be
		\frac{m_\omega^\ast}{m_\omega}\approx \kappa\frac{g_\omega^\ast}{g_\omega} = \kappa\Phi^\ast_\omega(n).
		\ee
		In the numerical calculation, we take
		\be
		\Phi^\ast_\omega\equiv \frac{g_\omega^\ast}{g_\omega}\approx 1-d\frac{n-n_{1/2}}{n_0}
		\ee
		with $d\approx 0.05$.
		\item Nucleon and dilaton masses: We simply regard that they are density invariant in this region, that is,  
		\be
		\frac{m_N^\ast}{m_N} & \approx & \frac{f_\chi^\ast}{f_\chi}\approx \frac{f_\pi^\ast}{f_\pi} \equiv \kappa\sim (0.6-0.9),\nonumber\\
		\frac{m_\sigma^\ast}{m_\sigma} & \approx & \kappa^{1+\beta^\prime/2}.
		\label{kappa}
		\ee
	
	\end{itemize}

\end{itemize}

We finally want to emphasize that the BR scaling which incorporates the density effect in the bare parameters in the effective Lagrangian which has the same form as in the vacuum is intrinsic density dependence (IDD). In addition to IDD, there is also an induced density effect (DD$_{\rm induced}$) arising from nucleon correlations~\cite{Paeng:2017qvp}. In the physical system, both effects should be considered.

\section{The quenched $g_A$ problem}

\label{sec:gA}

By using the above discussed b$s$HLS, we now show how the scale symmetry manifests in nuclei by looking at the quenched $g_A$ problem in the nuclear Gamow-Teller transitions, that is, the axial coupling constant $g_A^{\rm free} = 1.276$ in the vacuum should be replaced by an effective value $g_A^{\rm eff} \to 1 $ in the nuclear Gamow-Teller transitions~\cite{Wilkinson:1973zz,Suhonen:2017krv,Engel:2016xgb}. This "universal quenching" has been a mystery for over four decades, with explanations ranging from fundamental QCD renormalization to many-body correlations~\cite{Engel:2016xgb,Suhonen:2017krv}. The BR scaling anchored on the b$s$HLS gives a natural resolution and illustrates how the scale symmetry manifests at densities $\sim n_0$.

\subsection{Scale symmetry in the axial current}

In b$s$HLS~\eqref{eq:LbsHLS}, the axial current relevant to the nuclear Gamow-Teller transitions can be obtained through the external sources by gauging the chiral symmetry in Eq.~ \eqref{eq:1form}. It is finally expressed as
\be
q_{\rm SSB}g_A\bar{\psi}\tau^+\gamma_\mu\gamma_5 \psi
\label{eq:WeakCurrent}
\ee
where $\tau^+ = (\tau^1+i\tau^2)/2$ with $\tau^i$ being the Pauli matrices and
\be
q_{\rm SSB} = c_A + (1-c_A)\Phi^{\ast \beta^\prime},
\label{eq:qssb}
\ee
denotes the effect of the dilaton compensator. In matter-free space, $q_{\rm SSB}=1$ since $\Phi^\ast=1$. Therefore, the axial current coupling is intact, irrelevant to the parameter $c_A$ and the anomalous dimension of gluon field $\beta^\prime$.

The LOSS approximation ($c_A \approx 1$) implies that scale symmetry breaking does not affect the axial coupling—the fundamental $g_A$ remains unscaled with density. This is a crucial insight: the axial current operator is scale-invariant; only the nucleon mass term couples to the dilaton. This is precisely the BR scaling result: $g_A^\ast/g_A = 1$. However, when $c_A < 1$ which accounts for the breaking of LOSS and reflects the patterns of the scale symmetry in nuclei system, $q_{\rm SSB} < 1$ since $\Phi^\ast< 1$ in nuclear medium. In this case, the axial current coupling $g_A$ is modified by dilaton condensate.

Using \eqref{eq:WeakCurrent}, the total quench factor is finally expressed as~\cite{Ma:2020tsj}
\be
q_{\rm bsHLS} = q_{\rm SSB}\times q_{\rm SNC},
\label{eq:qGnEFT}
\ee
where $q_{\rm SNC}$ accounts for strong nuclear many-body correlations which can be calculated by using the Landau-Fermi fixed point theory.

\subsection{Landau Fermi-Liquid Fixed Point Theory}

The quench factor $q_{\rm SNC}$ can be calculated in terms of the Landau-Fermi fixed point theory~\cite{Friman:1996qc} using bsHLS. The key ingredient for this is that the
mean-field approximation with the bsHLS Lagrangian endowed
with the IDDs inherited from QCD together with DD$_{\rm induced}$ corresponds to Landau-Fermi liquid fixed point theory à la Wilsonian renormalization group to many-fermion
systems with Fermi surface~\cite{Paeng:2015noa}. 
Explicit calculation yields 
\be
q_{\rm SNC}^{\rm Landau} = \left(1-\frac{1}{3}\Phi^\ast \bar{F}_1^\pi\right)^{-2} ,
\label{eq:qSNC}
\ee
where $\Phi^\ast  \bar{F}_1^\pi$ is the pion Fock contribution to the Landau parameter. At normal nuclear matter density, from deeply bound pionic system~\cite{Kienle:2004hq}, we have
\be
\Phi^\ast (n_0) \simeq 0.8,
\ee
and~\cite{Brown:1980bg,Friman:1996qc}
\be
\bar{F}_1^\pi(n_0) = - 0.459.
\ee
Thus
\be
q_{\rm SNC}^{\rm Landau}(n_0) \approx 0.79~~~{\rm at}~n=n_0.
\label{eq:qSNCn}
\ee

It turns out that the product $\Phi^\ast \bar{F}_1^\pi$, and therefore \eqref{eq:qSNC} is nearly density-independent, so \eqref{eq:qSNCn} applies to both light and heavy nuclei. 
The BR scaling of $f_\pi^\ast$ (and hence $\Phi^\ast $) plays a central role in this result.

\subsection{The quench factor for light and heavy nuclei}

Now, let us consider the total quench factor with respect to data and some theoretical estimations of the matrix elements.  With respect to the experimental situation which will be specified later, we consider two nuclear mass regions: $A \lesssim 60$ (light) and $A > 60$ (heavy).

In nuclei up to $A\sim 60$, $g_A^{\rm eff}$ in shell model comes out to be~\cite{Suhonen:2017krv,Engel:2016xgb}
\be
g_A^{\rm eff} =q_{\rm light}g_A^{\rm free} = 0.98-1.18.
\ee
This implies the total quench factor $q_{\rm light}=0.76 - 0.93$. If $g_A^{\rm eff} \simeq 1$, we have
\be
q_{\rm light} \simeq 0.78.
\ee
With respect to \eqref{eq:qSNCn}, one concludes that $q_{\rm SSB}\simeq1$ from Eq.~\eqref{eq:qGnEFT}. This indicates that $c_A =1$ in Eq.~\eqref{eq:qssb},  which implies that LOSS works well for the light nuclei system.

However, as the mass number of nuclei goes up above $A \sim 60$, the situation is different. A more stringent recent experiment from RIKEN on the superallowed GT decay of the doubly magic nucleus $^{100}$Sn~\cite{Lubos:2019nik} yields the total quench factor~\cite{Ma:2020tsj}
\be
q_{\rm RIKEN}^{\rm ESPM} = 0.46-0.55,
\ee
which is reasonably obtained by the extreme single particle shell model (ESPM).
This means that in the heavy nuclei system, only~\eqref{eq:qSNCn} is not enough to account for the quench factor. A further suppression 
\be
q_{\rm SSB} \simeq 0.64
\ee
which comes from \eqref{eq:qGnEFT} is needed. This discrepancy can be interpreted by assuming $c_A \simeq 0.15$ and $\beta^\prime \simeq 2.5$---the same values that resolve the HWZ problem in Refs.~\cite{Ma:2016nki,Shao:2022njr}. In this case, 
\be
q^{\rm ESPM} = q_{\rm SSB} \times q_{\rm SNC} = 0.64 \times 0.79  \simeq 0.5
\ee
which agrees well with the RIKEN data.

This implies that the anomalous dimension of gluon field $\beta^\prime$ thus plays a crucial role in heavy nuclei, even though the LOSS approximation suffices for light nuclei and compact stars. This suggests a continuity: at low density (light nuclei), scale symmetry emerges (unitarity limit); at normal nuclear matter density, it is hidden; at high density (compact stars) and in heavy nuclei, it reappears through the anomalous dimension.

Moreover, we can conclude that the "fundamental" $g_A$ is unscaled with density ($g_A^\ast/g_A = 1$). The observed quenching is due to nuclear correlations (Landau Fermi-liquid effects) and, in heavy nuclei, anomalous dimension effects.

\section{The pseudoconformal structure of dense nuclear matter}

\label{sec:PCM}

Let us turn to the NM properties relevant to the cores of massive stars by using the b$s$HLS with BR scaling---terminologized as G$n$EFT. To take the meson fluctuation effects into account, we use the $V_{low k}$ approach~\cite{Bogner:2003wn}. Therefore, the density effects come from both the intrinsic density dependence (IDD) inherited from the BR scaling and nucleon correlations (DD$_{\rm induced}$).

Since in the cores of massive stars, the density could approach $\sim 10n_0$ with explicit value depending on model, the BR scaling is segmented to two regions by $n_{1/2}$, as we discussed above.  The density $n_{1/2}$ where the topology change happens and how the R-I and R-II are delineated changes density dependence of the hadron properties drastically and therefore impact the EoS in a qualitative way. However, as we discussed above, the location of the topology change is model dependent so that we cannot pin down its value theoretically. With respect to constraints from the heavy ion collisions~\cite{Tsang:2018kqj}, we constrain $n_{1/2}$ as $2.0n_0 \lesssim n_{1/2}\lesssim 4.0n_0$~\cite{Ma:2018xjw}.

\subsection{Pseudoconformal structure}

Before calculating the EoS of NM, let us first understand the implication of the conformal SV $v_s^2 = 1/3$ in NM~\cite{Ma:2023ugl}.

In NM, the SV is defined by
\be
v_s^2=\frac{\partial P(n)}{\partial n}/\frac{\partial\epsilon}{\partial n},
\ee
where $\epsilon$ and $P$ are, respectively, the energy density and the pressure. 
For a constant SV, after integration with respect to density, one obtains
\be
P(n) = v_s^2 \epsilon(n) - \frac{1}{3}C_0,
\ee
with $C_0$ being the integration constant. And for $v_s^2 = 1/3$---the conformal SV, we obtain
\be
\epsilon(n) - 3 P(n) = C_0.
\ee
For an ideal liquid system where the NM is assumed to work, one has
\be
\langle\theta_\mu^\mu\rangle = C_0,
\ee
which is a density independent quantity. 

When $C_0=0$, the TEMT vanishes so that the scale symmetry is exactly restored. The system with this property can be regarded as that where both the scale and chiral symmetries are restored, quarks are deconfined and perturbative QCD applies. Since this scenario happens at the density much beyond the cores of massive stars, we are not interested in it.

In the scenario $C_0 \neq 0$, that is, the TEMT is a density independent quantity, the conformal symmetry is not restored, although the SV saturates the conformal limit. The structure of the matter with these propertise is called pseudoconformal structure~\cite{Paeng:2017qvp,Ma:2018jze,Ma:2018xjw}. This scenario does happen in the chiral-scale EFT approach to NM. In the mean field approach it is shown that, going toward the DLFP, the TEMT $\langle\theta_\mu^\mu\rangle$ is a function of only the dilaton condensate~\cite{Paeng:2011hy}. Now if the condensate goes to a constant $\sim m_0$, the $\langle\theta_\mu^\mu\rangle$ will be independent of density. This chain of reasoning is confirmed in the full $V_{lowk}$ RG formalism as shown in Fig.~\ref{fig:TEMT} in the following.

\subsection{Equation of state and sound velocity}

We next explicitly compute the EoS of the pseudoconformal dense NM and compare it to the constraint from astrophysical observations and gravitational wave detections. In addition, we also vary the last parameter in the model, $n_{1/2}$, to see its effect on the EoS. We should say that, whatever the topology change density $n_{1/2} \gtrsim 2n_0$ is, the properties of ordinary NM are fixed. In addition, we assume and actually numerically checked that for $n \gtrsim n_{1/2}$, slightly above that transition density, the SV must be $v_s^2\approx 1/3$.

By varying the scaling parameters in R-II as that discussed in subsection~\ref{subsec:ParaBR}, it turns out that the energy per-particle at $n> n_{1/2}$ can be captured by a simple two-parameter formula:
\be
E/A= - m_N +X^\alpha  x^{1/3} + Y^\alpha x^{-1},
\label{PC-RII}
\ee
where $X$ and $Y$ are parameters to be fixed  by the model calculation, $\alpha=(N-Z)/(N+Z)$ being the asymmetry of neutrons and protons in NM and $x\equiv n/n_0$. From \eqref{PC-RII}, one concludes that the SV satisfies
\be
v_s^2=\frac{1}{3},
\label{pc-sound}
\ee
independently of $X^\alpha$ and $Y^\alpha$. Then the full EoS can be obtained by pin down the EoS in R-I and R-II with the parameters $X^\alpha$ and $Y^\alpha$ fixed by the continuity at $n=n_{1/2}$ of the chemical potential and pressure
\be
\mu_I=\mu_{II},\ P_I=P_{II}\ \ {\rm at} \ \ n=n_{1/2}.
\label{matchingE}
\ee

\begin{figure}
	\begin{center}
		\includegraphics[width=7.0cm]{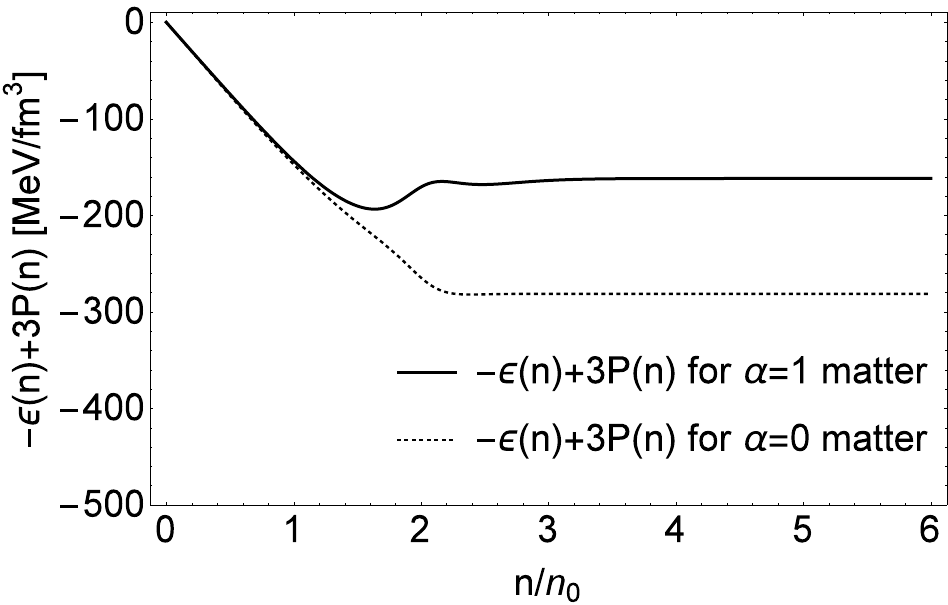}
		\includegraphics[width=7.0cm]{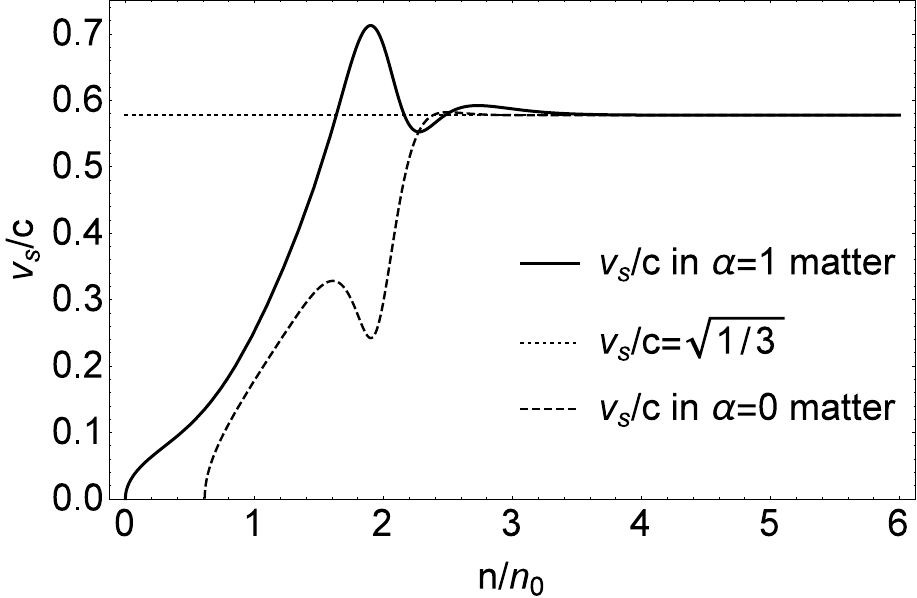}
		\caption{$\langle\theta_\mu^\mu\rangle$ (upper panel) and $v_s$ (lower panel) vs. density for $\alpha=0$ (nuclear matter) and $\alpha=1$ (neutron matter) in $V_{lowk}$ RG for $n_{1/2}=2 n_0$ and $v_{\rm vm}=25 n_0$.
		}\label{fig:TEMT}
	\end{center}
\end{figure}

We plot in Fig.~\ref{fig:TEMT} the SV and TEMT as functions of density. From this figure one can easily see that after $n_{1/2}$, the conformal SV $v_s^2=1/3$ emerges. The spike structure could very well be an artifact of the sharp connection made at the boundary. What is however physical is the rapid increase of the SV at the transition point signaling the changeover of degrees of freedom (DoFs)~\cite{Hippert:2021gfs} and the derivative contribution from the trace anomaly~\cite{Fujimoto:2022ohj,Marczenko:2022jhl}. 

The plot of the TEMT shows however that it is a nonzero constant, that is, matter is the "pseudoconformal SV" matter. The deviation from true conformal symmetry is encoded in the constant $C_0 = \langle \theta_\mu^\mu \rangle = 4n_0Y^\alpha$, where $Y^\alpha$ is fixed by the continuity at $n_{1/2}$. This is the key insight: the TEMT is not zero (scale symmetry is still broken), but it is density-independent, leading to the pseudoconformal SV.

\begin{figure}
	\begin{center}
		\includegraphics[width=7cm]{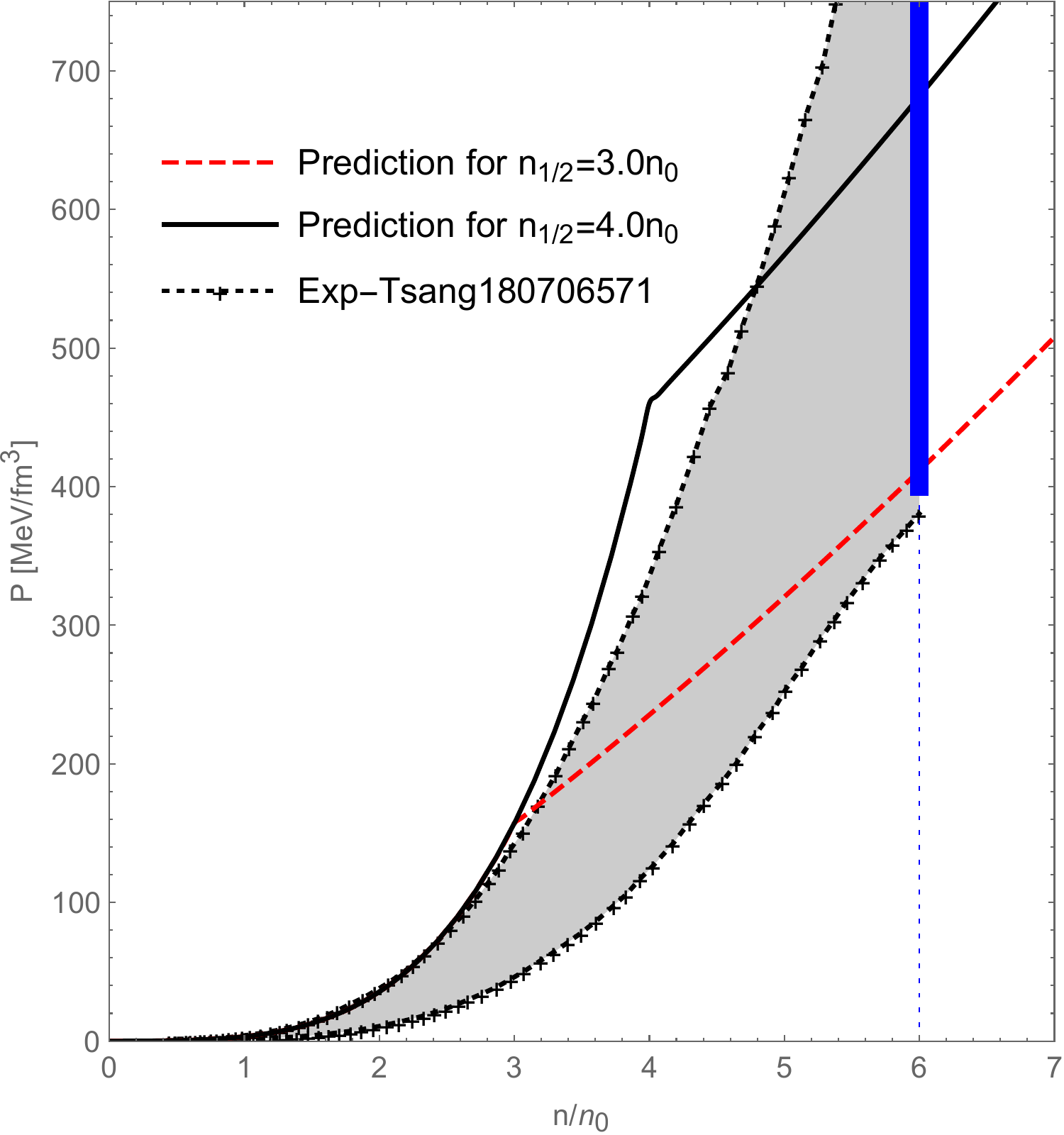}
	\end{center}
	\caption{Predicted pressure for neutron matter ($\alpha=1$) vs density compared  with the available experimental bound (shaded) given by Ref.~\cite{Tsang:2018kqj} and the bound at $6n_0$ given by the blue band.}
	\label{Tsang}
\end{figure}

Plotted in Fig.~\ref{Tsang} is the predicted pressure $P$ for $n_{1/2}/n_0= 3,4$ compared with the presently available heavy-ion data~\cite{Tsang:2018kqj}.
The case of $n_{1/2}=4 n_0$, while consistent with the bound at $n\sim 6n_0$,  goes outside of the  presently available experimental bound at $n\sim 4n_0$. In combination with the analysis for $n_{1/2} = 2n_0$, this  allows us to set the constraint for $n_{1/2}$
\be
2.0 n_0 \lesssim  n_{1/2} \lesssim  4.0 n_0.\label{boundhalf}
\ee
What is important about this constraint is that the emergence of the conformal SV is an order of magnitude lower than the asymptotic density $\gtrsim 50 n_0$ perturbative QCD predicts. This signals the precocious emergence of pseudo-conformality in compact stars. A recent detailed analysis of currently available data in the quarkyonic model does confirm the onset density of $v_s^2\approx 1/3$ at $\sim 4 n_0$~\cite{Zhao:2020dvu,Kapusta:2021ney,Margueron:2021dtx}

\subsection{Comparison with observations}

To validate the pseudoconformal model (PCM) of dense NM, we present the predictions of a variety of neutron star properties. We should say that since the normal NM is in R-I, its properties constrain the parameters in $n<n_{1/2}$ and, therefore indirectly affect the EoS in R-II through the matching condition \eqref{matchingE}. The predicted neutron star properties are summarized as~\cite{Paeng:2017qvp,Ma:2018jze,Ma:2018xjw}:
\begin{enumerate}
	\item Maximum neutron star mass: $2.04 M_\odot \lesssim M_{\rm max} \lesssim 2.23 M_\odot$ for $2.0 \le n_{1/2}/n_0 \le 4.0 $.
	
	\item Tidal deformability for $1.4M_\odot$ is $\Lambda_{1.4} \approx 650$, consistent with the LIGO/Virgo constraint $\Lambda_{1.4} < 800$.
	
	\item Radius for $1.4M_\odot$: $R_{1.4} \approx 12.8$~km.
	
\end{enumerate}
All these predictions are consistent with the observations. Therefore, the PCM of dense NM cannot be ruled out by the present data.

The pseudoconformal SV can be intimately connected to the parity-doubling structure of the nucleon which allows a chiral invariant mass of nucleon $m_0$~\cite{Detar:1988kn,Jido:2001nt} that survives after the topology change~\cite{Ma:2013ooa}. Since $m_N^\ast \to m_0$ (density-independent), the dilaton condensate, which controls the nucleon mass, is constant. This constancy propagates through the trace anomaly to yield $\langle \theta_\mu^\mu \rangle=$constant. The magnitude of $m_0$ determines the size of $C_0$:
\be
C_0 \sim m_0^4
\ee
indicating that the chiral-invariant mass component $m_0$ is the source of the nonzero TEMT in the pseudo-conformal phase. This provides a direct link between the origin of nucleon mass and the macroscopic properties of compact stars. The constraints from the compact star observations yields $m_0 \approx (0.5-0.9) m_N$ then can estimate the PCM constant $C_0$~\cite{Gao:2024lzu,Gao:2025nkg,Gao:2025vdc}. 

\section{Peak of sound velocity}

\label{sec:peak}

In astrophysics, the observations of two-solar mass neutron stars~\cite{Demorest:2010bx,Antoniadis:2013pzd} imply that the EoS should be stiff at high densities while the measurements of the radii of intermediate mass neutron stars~\cite{Vinciguerra:2023qxq,Choudhury:2024xbk,Mauviard:2025dmd} indicate that the EoS is soft at intermediate densities. These combined information suggests that the SV favors a peak structure at intermediate density~\cite{Altiparmak:2022bke,Marczenko:2022jhl}. In literature, the microscopic origin of this peak is normally attributed to the transition from hadronic matter to exotic configurations or emergence of new DoFs~\cite{McLerran:2018hbz,Jeong:2019lhv,Zhao:2020dvu,Margueron:2021dtx,Ma:2018xjw,Ma:2019ery,Ma:2021nuf,Hippert:2021gfs,Alford:2004pf,Baym:2019iky,Chiba:2023ftg,Alford:2017qgh,Leonhardt:2019fua}.

I next will show that the peak emerges automatically from b$s$HLS~\eqref{eq:LbsHLS} with the standard RMF approach without referring to the phase or configuration change in the matter~\cite{Zhang:2024sju,Zhang:2024iye}. This novel phenomenon is attributed to the nonlinear realization of scale symmetry regarded as the source of scalar meson, which is different from Walecka-type quantum hadron dynamical models.  Additionally, to restore scale symmetry in dense matter, the effective $\omega$-$N$-$N$ coupling decreases faster than $\rho$-$N$-$N$ in-medium, thereby breaking the $\rm U(2)$ flavor symmetry and LOSS, which works well in the matter-free space. These observations give another hint on how the scale symmetry manifests in NM.

\subsection{Parameter estimation}

In the calculation, to saturate the constraints from symmetric NM properties around saturation density, we include the medium effect on the bare parameters in the Lagrangian using Brown-Rho scaling~\eqref{eq:BRScaling} and \eqref{eq:BRScalingsigma}, that is, the IDD. After solving the equations of motion of hadron fields, an induced density dependence---DD$_{\rm induced}$---arises from nucleon correlations, e.g. number density \(\langle N^\dagger N\rangle^*\) and scalar density \(\langle \bar{N}N\rangle^*\). The full density dependence of the system is a sum of both~\cite{Ma:2019ery}.

In the RMF calculation, the results depend on the following combinations of the parameters:
\be
& & g_{\omega NN}=\frac{1}{2}\left(g_{V_\rho} C_{V_\rho}+g_{V_0} C_{V_0}-1\right) g_\omega\ ,\nonumber\\
& & g_{\rho NN}=\frac{1}{2}\left(g_{V_\rho} C_{V_\rho}-1\right) g_\rho\ ,\nonumber\\
& & g^{SSB}_{\omega NN}=\frac{1}{2}\left[g_{V_\rho}\left(1-C_{V_\rho}\right)+g_{V_0}\left(1-C_{V_0}\right)\right]g_\omega\ ,\nonumber\\
& & g^{SSB}_{\rho NN}=\frac{1}{2}\left[g_{V_\rho}\left(1-C_{V_\rho}\right)\right]g_\rho\ .
\ee

\begin{table}[htbp]\small
	\centering
	\caption{
		Values of the saturation density $n_0$, binding energy $e_0$, symmetry energy $E_{\mathrm{sym}}$, incompressibility coefficient $K$, skewness coefficient $J$ and symmetry energy density slope $L$. $n_0$ is in unit of $\mathrm{fm^{-3}}$ and others are in units of $\mathrm{MeV}$.
	}
	\begin{threeparttable}~\label{tab:nuclear-matter}
		\begin{tabular}{@{}cccc}
			\hline
			\hline
			& Empirical & bsHLS-L & bsHLS-H \\
			\hline
			$n_0$ & $0.155\pm0.050$~\cite{Sedrakian:2022ata} & 0.159 & 0.159 \\
			\hline
			$e_0$ & $-15.0\pm1.0$~\cite{Sedrakian:2022ata} & $-16.0$ & $-16.0$ \\
			\hline
			$K(n_0)$ & $230\pm30$~\cite{Dutra:2012mb} & 232 & 284\\
			\hline
			$E_{\mathrm{sym}}(n_0)$ & $30.9\pm1.9$~\cite{Lattimer:2012xj} & 30.5 & 29.2\\
			\hline
			$E_{\mathrm{sym}}(2n_0)$ & $46.9\pm10.1$~\cite{Li:2019xxz} & 51.5 & 50.2\\
			\hline
			$L(n_0)$ & $52.5\pm17.5$~\cite{Lattimer:2012xj} & 85.9 & 68.3 \\
			\hline
			$J(n_0)$ & $-700\pm500$~\cite{Farine:1997vuz} & $-767$ & $-599$\\
			\hline
			\hline
		\end{tabular}
	\end{threeparttable}
\end{table}

In the calculation, we use the NM properties as constraints to estimate the parameters. We find that, as shown in Tab.~\ref{tab:nuclear-matter}, the symmetric NM properties can be well reproduced using the parameters list in Tab.~\ref{tab:para}.
Among the possible sets of parameters, we list two of them, bsHLS-L and bsHLS-H, in Tab.~\ref{tab:para}. Although both bsHLS-L and bsHLS-H can give reasonable predictions, considering that pion-nuclei bound state data~\cite{Kienle:2004hq}, which gives $r \simeq 0.2$ and the skyrmion crystal estimation~ $1.0 \lesssim |\gamma_{G^2}| \lesssim 3.5$~\cite{Ma:2016nki,Shao:2022njr}, bsHLS-H is preferred.
Moreover, using bsHLS-H, we obtain $K_{\text {sym }}=9 n_0^2 \frac{\partial^2 E_{\text {sym }}}{\partial n^2}|_{n=n_0}=-54.0\ \mathrm{MeV}$ which yields \(K_\tau \approx K_{\text {sym }}-6 L(n_0)=-464\ \mathrm{MeV}\), consistent with the constraint \(K_\tau \approx-550 \pm 100~\mathrm{MeV}\)~\cite{Centelles:2008vu}. Actually, if one further considers the data of neutron star properties and tidal deformability from gravitational wave detection~\cite{LIGOScientific:2018cki,LIGOScientific:2018hze,Riley:2019yda,Riley:2021pdl}, the rationality of bsHLS-H is reinforced~\cite{Zhang:2024iye}.

{\small
\begin{table}[htbp]\small
	\centering
	\caption{
		Values of parameters for bsHLS-L and bsHLS-H obtained from fitting the NM properties in Tab.~\ref{tab:nuclear-matter}. $M_\sigma = f_\chi m_\sigma$ is in units of $10^5\rm MeV^2$.
	}
	\begin{threeparttable}
		~\label{tab:para}
		\begin{tabular}{@{}cccccccc}
			\hline
			\hline
			& \(M_{\sigma}\) & \(\beta'\) & \(r\) & \(g_{\omega NN}\) & \(g_{\rho NN}\) & \(g_{\omega NN}^{SSB}\) & \(g_{\rho NN}^{SSB}\)\\
			\hline
			bsHLS-L & 1.05 & 0.395 & 0.161 & 11.5 & 3.78 & 16.3 & 9.45\\
			\hline
			bsHLS-H & 2.30 & 1.15 & 0.191 & 11.0 & 4.17 & 8.85 & 4.85\\
			\hline
			\hline
		\end{tabular}
	\end{threeparttable}
\end{table}
}

\subsection{Sound velocity}

\begin{figure}
	\centering
		\includegraphics[width=0.455\textwidth]{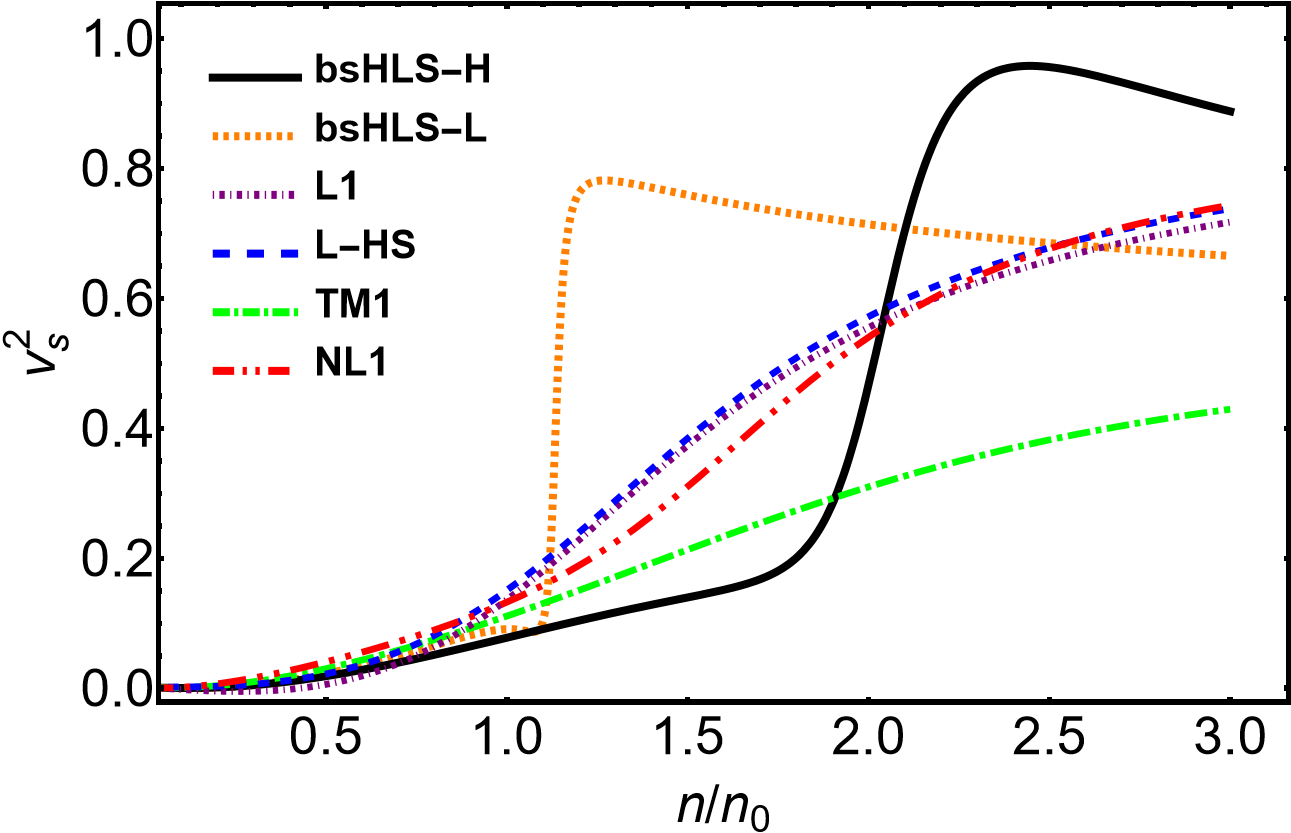}
	\caption{The speed of sound in PNM for various models. The results from the choices of parameters in  Walecka-type models L1~\cite{Lee:1986bi}, L-HS~\cite{Horowitz:1981xw}, NL1~\cite{Reinhard:1986qq} and TM1~\cite{Sugahara:1993wz} are shown for comparison. }~\label{fig:sound-bsHLS}
\end{figure}

The results of SV in pure neutron matter (PNM) from the two parameter sets are plotted in Fig.~\ref{fig:sound-bsHLS}. One can see that, there is indeed a peak behavior which locates at intermediate densities around $(1 - 2.5)\ n_0$.
In contrast to the above referred studies where the peak is attributed to the hadron-quark transition or configuration change, here, only hadron DoFs exist and no new configuration emerges. In addition, since we used the uniform BR scaling~\eqref{eq:BRScaling} and \eqref{eq:BRScalingsigma} without considering the difference of the density dependence before and after $n_{1/2}$ discussed above, the peak obtained here is not due to the topology change discussed in last section. Furthermore, this peak is not found in the Walecka-type models~\cite{Lee:1986bi,Horowitz:1981xw, Reinhard:1986qq, Sugahara:1993wz} as compared in Fig.~\ref{fig:sound-bsHLS}. Therefore, this unique feature, not been observed in other hadronic models, emerges from the realization of the trace anomaly of QCD at hadronic level, as will be discussed in the following.

\begin{figure}
	\centering
	\includegraphics[width=0.455\textwidth]{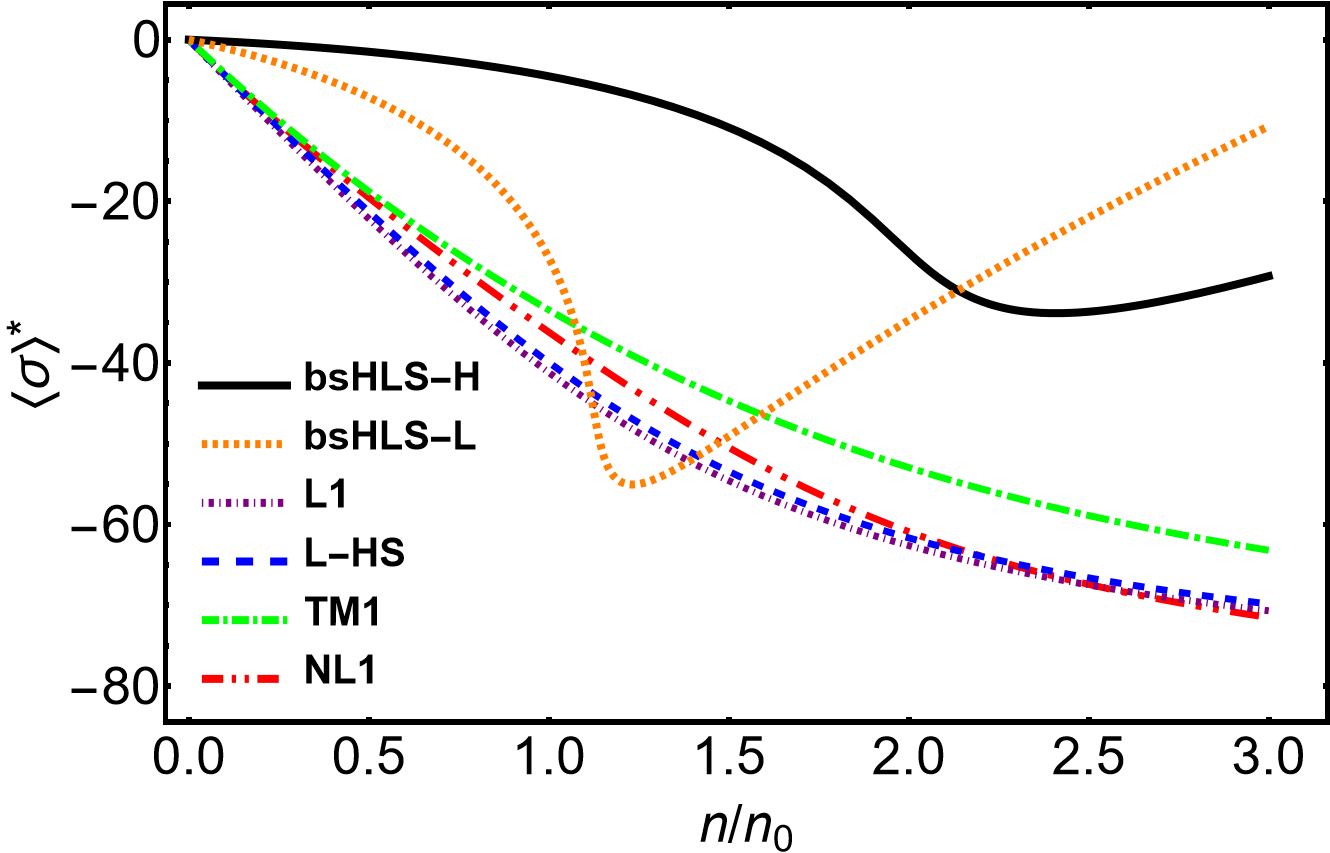}
	\caption{
		The mean field of $\sigma$ field (in unit of MeV) in PNM.
		\(f_\chi\) is chosen as \(f_\chi\approx f_\pi=92.4\ {\rm MeV}\) to keep consistency of chiral expansion~\cite{Crewther:2013vea}.
		Notations are the same as in Fig.~\ref{fig:sound-bsHLS}.
	}~\label{fig:sigma}
\end{figure}

Actually, this peak is locked to the solution of the dilaton field in b$s$HLS. In Fig.~\ref{fig:sigma}, we plot the mean field $\sigma$ in the dilaton compensator $\chi = f_\chi e^{\sigma/f_\chi}$ and that in the widely used Walecka-type models~\cite{Lee:1986bi,Horowitz:1981xw, Reinhard:1986qq, Sugahara:1993wz}. It can be seen that  \(\sigma\) in all the Walecka-type models have a monotonous density dependence, but there are kink behaviors in both bsHLS-L and bsHLS-H. And, the location of the kink is bounded to that of the peak. This numerical results confirm that the peak in $v_s^2$ is due to the kink behavior of $\sigma$ field in bsHLS-L and bsHLS-H---\(\sigma\) is nonlinearly coupled with other hadron fields through the conformal compensator---which is different from Walecka-type models.
This is the key factor that the bsHLS $\sigma$ field can generate the $v_s$ peak without resorting to any transition of phase or configuration.

It is interesting to note that the peak occurs at $n \sim (1-2.5)n_0$, a density range accessible to terrestrial experiments such as heavy-ion collisions at facilities like HIAF, RAON, RHIC, and FRIB. So that, whether this picture is physically accepted are promising to be tested.

\subsection{Scale symmetry in medium}

The above discussion suggests that the peak of the SV emerges from the dilaton compensator approach. Then, what is the conformality in NM? 
Upon closer inspection of Fig.~\ref{fig:sigma}, one may observe that the magnitude of spontaneous scale symmetry breaking increases with density after the kink, contrary to naive expectations, because the order parameter of scale symmetry---\(\langle\chi\rangle^*\) increasing with \(\langle\sigma\rangle^*\).
Actually, this phenomenon is also found in skyrmion crystal approach to dense matter using the dilaton compensated EFT~\cite{Park:2003sd,Park:2008zg,Ma:2013ela,Shao:2022njr}. Therefore, it seems that this behaviour is generic in the dilaton compensator approach. The density behavior can be understood by noting that in the dilaton compensator approach, the vector meson mass scales with density, e.g., $m_\omega^\ast = \Phi^\ast m_\omega$.
Then, if the omega mass were to decrease continuously as required by the VM~\cite{Harada:2000kb}, the repulsive omega force would increase with density. This force the omega meson mass $m_\omega^\ast$, hence $\Phi$, to increase and, consequently, the magnitude of the scale symmetry breaking increases~\cite{Park:2008zg}.

\begin{figure}
	\centering
	\includegraphics[width=0.455\textwidth]{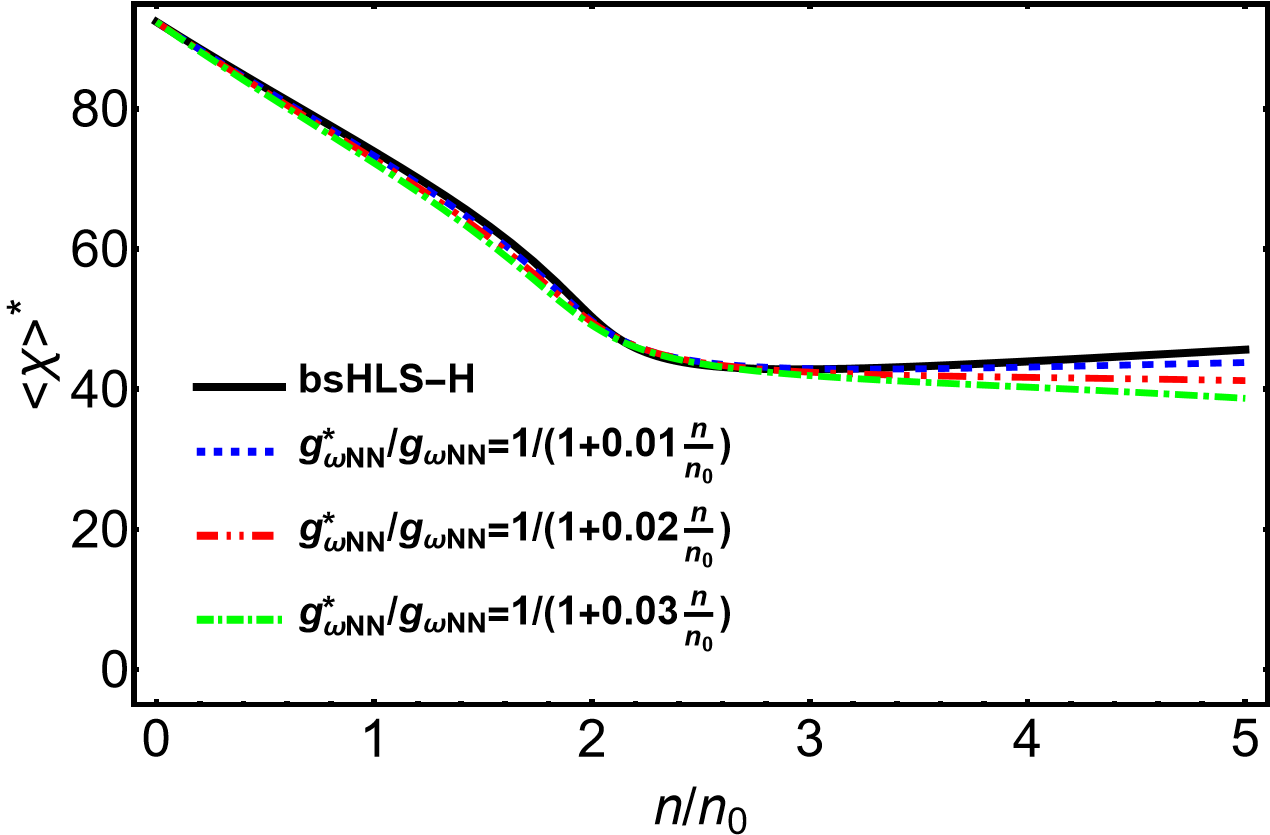}
	\caption{
		The expectation value of $\chi$ as a function of density with modified $\omega$-$N$-$N$ coupling, where \(f_{\chi}\) is set equal to \(f_{\pi}\).
	}~\label{fig:VEVChieff}
\end{figure}

\begin{figure}
	\centering
	\includegraphics[width=0.455\textwidth]{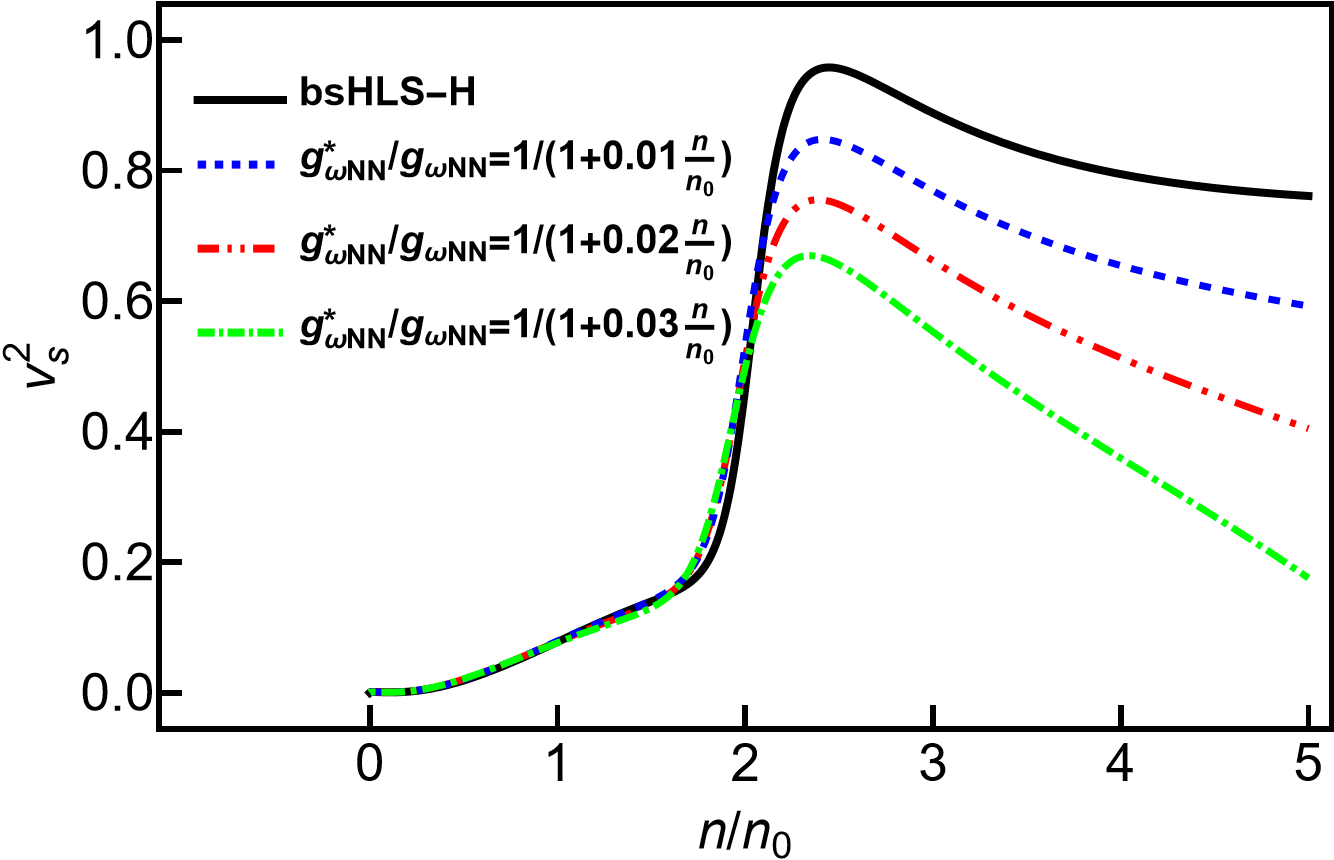}
	\caption{
		The sound velocity in PNM with modified $\omega$-$N$-$N$ coupling.
	}~\label{fig:sound velocityeff}
\end{figure}

One of the resolutions of this flaw is to couple more dilaton fields to the $\omega$-$N$-$N$ coupling to impose stronger density scaling and, consequently, suppress the effective coupling and therefore the repulsive force ~\cite{Park:2008zg,Paeng:2013xya,Shao:2022njr}. This can be numerical checked by typically choose factor $g_{\omega NN}^\ast/g_{\omega NN} \equiv 1/(1+ Rn/n_0)$ with $R \approx 0.02$. The VEV of dilaton field $\chi$ plotted in Fig.~\ref{fig:VEVChieff} clearly shows that, with this further suppression, the magnitude of scale symmetry breaking after the kink is reduced.
One may find that this suppression, especially the nearly density independence of $\langle \chi \rangle^\ast$, is very similar to the pseudo-conformality based on the topology consideration~\cite{Ma:2018xjw, Ma:2019ery, Ma:2021nuf, Lee:2021hrw, Ma:2023ugl}. At this moment, we cannot clearly build a connection between these two approaches. One possible way might be to look at the baryon number density distribution in the dilaton compensated chiral-quark model~\cite{MantovaniSarti:2013aqe} or  the baryon charge distribution using the Kohn-Sham theory~\cite{Hu:2019}.
An interesting observation is that the peak of SV remains, and its location is closely tied to the kink of $\langle \chi\rangle^\ast$, as illustrated in Fig.~\ref{fig:sound velocityeff}. One may find that the SV in Fig.~\ref{fig:sound velocityeff} does not reach consistently the asymptotic limit of QCD if our results are naively extrapolated to higher density. This is understandable since our approach only includes hadronic DoFs and such extrapolation is not solid.

In conclusion, the SV of NM---a fundamental observable characterizing the stiffness of the EoS---is extremely sensitive to the realization of scale symmetry and the BR scaling that follows from it. In the dilaton-compensated effective theory, the SV develops a characteristic peak at intermediate densities $n\sim (1-2.5)n_0$, arising from the nonlinear coupling of the dilaton field to hadron masses. This peak structure, absent in conventional models, provides a distinctive signature of scale symmetry in dense matter and has profound implications for neutron star structure and gravitational wave observations.

\section{Summary and discussion}

\label{sec:sum}

This review has traced the evolution of scale symmetry in dense baryonic matter and its profound implications for quenched $g_A$ problem in Gamow-Teller transition, the pseudoconformal structure of NM, the SV of NM. The genuine dilaton scenario, anchored on the Crewther-Tunstall infrared fixed point, provides a unified framework. The main findings can be summarized as follows:
\begin{enumerate}

	\item The Brown-Rho scaling relation finds its natural theoretical foundation in the scale-symmetric effective theory. The generalized BR scaling $m_N^\ast/m_N \simeq m_V^\ast/m_V \simeq f_\pi^\ast/f_\pi \simeq \Phi^\ast(n)$
    emerges from the coupling of hadrons to the dilaton condensate, with the dilaton mass scaling $m_\sigma^\ast/m_\sigma = \Phi^\ast(n)^{1+\beta^\prime/2}$ reflecting the explicit breaking of scale symmetry encoded in the anomalous dimension $\beta^\prime$.

	\item The quenched $g_A$ problem finds a natural resolution: the observed quenching is due to nuclear correlations (Landau Fermi-liquid effects) rather than fundamental QCD renormalization, with the anomalous dimension $\beta^\prime$ accounting for additional quenching in heavy nuclei. The fundamental $g_A$ remains unscaled with density—a direct consequence of the BR scaling of the axial current.

	\item Scale symmetry evolution follows three regimes: emergence at the unitarity limit (low density), hiding at normal nuclear matter density, and re-emergence at the dilaton limit (high density) with pseudo-conformal SV $v_s^2 = 1/3$. This continuity suggests that the scale symmetry emerging from nuclear correlations at high density is the same symmetry that manifests in the unitarity limit at low density, providing a unified picture of scale symmetry in nuclear dynamics across all density regimes.

\item The SV of NM exhibits a characteristic peak at intermediate densities $n \sim (1-2.5)n_0$, arising from the kink behavior of the dilaton field. This peak is a unique signature of scale symmetry realization that distinguishes the dilaton-compensated effective theory from conventional Walecka-type models.

\end{enumerate}

The framework presented here suggests a paradigm shift: nucleon mass is not solely due to spontaneous chiral symmetry breaking but receives a substantial contribution from emergent scale symmetry in dense matter. The SV peak predicted by the dilaton-compensated theory provides a unique experimental signature of this paradigm, accessible to heavy-ion collision experiments at intermediate densities. The BR scaling relation, once a phenomenological ansatz, is now understood as a fundamental consequence of scale symmetry in dense matter—a testament to the unifying power of scale symmetry in nuclear physics.

Although the above main achievements are yielded, there are still several open questions remain: (1) what is the relationship between the genuine dilaton and emergent scale symmetry, (2) whether the b$s$HLS developed here is consistent with the hadron processes, especially those including the scalar resonance and, (3) is it possible to have new BR scaling parameterizations that consistent with all the experiments and observations. Future nuclear physics experiments, neutron star observations and gravitational wave detections will impose the test of these progresses.

\section*{Acknowledgments}

I would like to express my deep gratitude for the collaborations and numerous valuable discussions with Mannque Rho since 2011.

Content presented in some parts of this contribution were borrowed from the collaborations with Mannque Rho and other co-authored colleagues for which I am very grateful.

This work is supported by the National Science Foundation of China (NSFC) under Grant No. 12547104 and Gusu Talent Innovation Program under Grant No. ZXL2024363.



\printcredits

\bibliographystyle{cas-model2-names}

\bibliography{cas-refs}





\end{document}